\documentclass{article}

\usepackage{amsmath,amsthm,amsfonts}

\usepackage{amssymb}

\usepackage{epsfig}

\usepackage{rotating}

\usepackage{subfigure}

\begin{document}

\centerline{\bf On the existence and scaling of structure functions in turbulence according to the data}
\vskip 12pt

\centerline{\bf Alexandre Arenas}
\centerline{Departament d'Enginyeria Inform{\`a}tica i Matem{\`a}tiques}
\centerline{Universitat Rovira i Virgili, 43007 Tarragona, Spain}
\centerline{and}
\centerline{\bf Alexandre J.\ Chorin}
\centerline{Department of Mathematics}
\centerline{University of California and Lawrence Berkeley National Laboratory}
\centerline{Berkeley, CA 94720}

%\affiliation{Departament d'Enginyeria Inform{\`a}tica i
%Matem{\`a}tiques, Universitat Rovira i Virgili, 43007 Tarragona, Spain}

%\author{Alexandre J. Chorin} \affiliation{Department of Mathematics,
%University of California, Berkeley, California 94720}

%\date{}

\baselineskip24pt
%\maketitle

\noindent
\begin{abstract}
We sample a velocity field that has an inertial spectrum and a skewness that matches experimental data. In particular, we compute a self-consistent
correction to the Kolmogorov exponent and find that for our model it is zero. We find that the higher order structure functions diverge for orders larger than a certain threshold, as theorized in some recent work. The significance of the results for the statistical theory of homogeneous turbulence is reviewed.
\end{abstract}

\section{Introduction}

In 1941 Kolmogorov \cite{kolm} formulated his famous scaling theory of the inertial range in turbulence,
according to which the second order structure function, i.e., the function
$S_2({\bf r})=<((u({\bf x + r})-u({\bf x}))^2>$, scales as $(\epsilon r)^{2/3}$, where $\bf x$, $\bf x+r$
are points in a turbulent flow field, $u$ is the component of the velocity in the direction of $\bf r$, $\epsilon$
is the mean rate of energy dissipation, $r$ is the length $|\bf r|$ of $\bf r$, and the brackets denote an average. A Fourier
transform yields the Kolmogorov-Obukhov inertial range spectrum $E(k)=C\epsilon^{2/3} k^{-5/3}$, where $C$ is a constant and $k$ is the wave number \cite{obukhov}.
These results are the lynchpins of turbulence theory yet uncertainty lingers as to their general validity and as to the details of the derivation.
Subsequently ( see e.g. \cite{frisch}) it was claimed that this 
scaling result generalizes to structure functions of any order, i.e., to 
$S_p({\bf r})=<((u({\bf x + r})-u({\bf x}))^p>$=$(\epsilon r)^{p/3}$ for any $p>0$, but note that Kolmogorov and Obukhov themselves
never went that far; Kolmogorov in \cite{kolm3} gave only the further result for $p=3$. 

Shortly after the publication of this theory, Landau \cite{landau} challenged its derivation on
the ground that the rate of energy dissipation is intermittent, i.e., spatially
inhomogenous, and cannot be treated as a constant. This observation fits in with the
experimental observation that for $p>3$ the exponents are smaller than the
values given by the extended Kolmogorov theory and, and it is also often claimed
for the $p=2$ the experimental value of the Kolmogorov-Obukhov exponent is larger than the prediction of this scaling theory. 
Kolmogorov and Obukhov themselves  produced a ``corrected" theory \cite{obukhov2}, \cite{kolm2}, and the current
belief is that the theory has to be supplemented by ``intermittency corrections";
much effort has been expended on the calculation of these corrections (see e.g. \cite{monin, frisch}).

A different analysis of possible corrections to the Kolmogorov exponents has
been offered  by Barenblatt and Chorin in \cite{bc1, bc2}. The structure functions depend on
the Reynolds number $R$ (for a definition of $R$ suitable for the present case
see \cite{b2}), a bulk length scale $L$, and the mean rate of energy dissipation $\epsilon$.
Dimensional analysis yields $S_p=(\epsilon r)^{p/3}\Phi_p(r/L,R)$, where $\Phi_p$
is an unknown dimensionless function of two large arguments. If one makes a complete
similarity assumption (see \cite{b3}), one finds $\Phi_p\sim C$ for large arguments, and one recovers the Kolmogorov
exponents above, 
but other assumptions are possible and indeed natural. In particular, an analogy
with a successful scaling theory for turbulent boundary layers \cite{bc4},
\cite{bc3}, leads to the assumption $\Phi_p=C_p(R)(r/L)^{(\alpha_p/lnR)}$, where $\alpha_p$ is a constant
and $C_p(R)$ is a function of the Reynolds number $R$. This yields structure functions $S_p$ proportional to $r^{p/3+\alpha_p/lnR}$, i.e., exponents that depend
on $R$, and maybe even more important, a proportionality constant that depends on $R$ and
may well diverge as $R \rightarrow \infty$ for $p$ large enough. As pointed out in \cite{bc4}, the dependence
on $1/lnR$ is an assumption, and in reality may be weaker yet; such dependence
could be hard to detect in experimental data.

This analysis is quite compatible with the Landau observation regarding intermittency,
as was already pointed out in \cite{bc3}: Turbulent flow is intrinsically intermittent, and indeed in the
limit of infinite $R$, extremely intermittent, in the sense that the bulk of the vorticity
occupies a negligible fraction of the available volume \cite{chorinbook}; this is a dominant fact about turbulence, and the notion of ``intermittency corrections" is no more meaningful than a
notion of ``turbulence corrections" to turbulent flow. However, at finite $R$ viscosity
reduces the intermittency and the scaling has to be corrected for ``intermittency reductions"
that depend on $R$.
In related work \cite{bc5}, Barenblatt and Chorin made numerical estimates on the basis
of an approximate theory of turbulence and came to the conclusion that, as $R$ increases, 
the structure functions for $p\ge4$ diverge; similar conclusions
had been reached by P.  Constantin (personal communication)  through mathematical considerations and by Mandelbrot \cite{mandel} by a geometrical similarity argument. Thus indeed $C_p(R)$ would diverge as $R$ increases for $p\ge4$.

In the absence of solutions of the Navier-Stokes equations these various theories
can only be checked by data from experiments, and this seems to be difficult at present. It occurred to us to interrogate further the available data about the inertial range
by building a stochastic computer model of the inertial range, compatible with well-accepted
and reproducible data about their skewness (indicating the level of intermittency),
and then examine the resulting structure functions. The field we construct is
obviously not unique;
however, any model that satisfies constraints imposed by the
data is highly instructive, and indicates  what possibilities are worthy of further study.
The results are striking enough to open new vistas for
theory, as we discuss in the concluding section.

We thus construct on the computer a one-dimensional Gaussian velocity field
with a power-law, then modify it so that the velocity differences assume the
skewnesses presented in Batchelor's book \cite{batch} on the basis of the data in \cite{stewart}; the power law will be either the Kolmogorov-Obukhov spectrum or a modification
that satisfies other constraints (see below).
We use a one-dimensional model so that we can perform the calculations
with sufficent accuracy, and think that this model is sufficient to carry our
conclusions. We then 
calculate, when possible, the higher-order structure functions. The data in \cite{batch, stewart} are tabulated there for various Reynolds numbers; we take the
data that correspond to the largest $R=42200$; our model has a power law for all
wave numbers and is thus inviscid. Note that by modern standards this value of $R$ is not very large; we thus assume 
that the skewness does not vary drastically as $R$ increases further. We first present the technical aspects of the construction, then the results, then provide a discussion.

\section{Building a Model}
We first explain how to sample effectively a homogenous Gaussian velocity field with a given spectrum. Our basic tool is a construction due to Elliott et al. \cite{elliott, elliott2}.  A homogeneous Gaussian random field is determined by its second order means:

\begin{align}
\nonumber <u(x)>=0, \\
<u(x+\Delta x)u(x)>=\int_{- \infty}^{\infty} e^{-2 \pi i k x} E(k) dk.
\label{correl}
\end{align}
The spectral representation of the Gaussian random field $u$ is

\begin{equation}
u(x)=\int_{- \infty}^{\infty} e^{2 \pi i k x} E^{1/2}(k) dw(k),
\label{spectral}
\end{equation}
\noindent where $w$ is a a Wiener process. Expand $dw$ in a complete orthonormal
series $\phi_m$:

\begin{equation}
dw(k) = \sum_{m=1}^{\infty} \gamma_m \phi_m(k) dk,
\label{sum}
\end{equation}
where the ${\gamma_m}$ form a sequence of independent random Gaussian variables. By substituting (\ref{sum}) into Eq. (\ref{spectral}) we find:

\begin{equation}
u(x)=\sum_{m} \gamma_m c_m(x),
\label{spectralsum}
\end{equation}

\noindent where the coefficients $c_m(x)$ are:

\begin{equation}
c_m(x)=\int_{- \infty}^{\infty} e^{2 \pi i k x} E^{1/2}(k) \phi_m(k) dk \equiv {\cal F}^{-1}[E^{1/2}\phi_m](x),
\label{coefficients}
\end{equation}
and ${\cal F}$ is the Fourier transform. Elliott et al. proposed to use as basis of the decomposition, Fourier transforms of  wavelets based on the Meyer wavelet; these wavelets are  generated from the Meyer mother wavelet $\psi(k)$ by the wavelet relation:

\begin{equation}
\psi_{mn} (x)= 2^{-m/2} \psi (2^{-m}x-n),
\label{kernelspectrum}
\end{equation}

\noindent where the index $m$ refers to different scales (octaves) and $n$ to different dilations. Using this prescription, Eq. (\ref{spectralsum}) can be represented as,

\begin{align}
\nonumber u(x)=\sum_m \sum_n \gamma_{mn} c_{mn}(x) \\
c_{mn}(x)={\cal F}^{-1} [E_{m}^{1/2} \phi](2^{-m}x -n),
\label{spectralmeyer}
\end{align}

\noindent where $E_{m}^{1/2}(k)=2^{-m/2}E^{1/2}(2^{-m}k)$. In the particular case of the
Kolmogorov-Obukhov energy spectrum $E(k) \sim k^{-5/3}$, the Fourier transform of the $c_{mn}$ coefficients represented in Eq. (\ref {spectralmeyer}) becomes:

\begin{equation}
({\cal F }c_{mn})(k)= 2^{-m/2} {\mid 2^{-m/2}k \mid}^{-5/6} \psi_{mn}(k).
\label{Kcoef}
\end{equation}
Similar formulas are obtained for every power-law spectrum. \newline Defining $f_{m}(2^{-m}x-n)= {\cal F}^{-1} (2^{-m/2} {\mid 2^{-m/2}k \mid}^{-5/6} \psi_{mn}(k))$, the representation of the field given in Eq.(\ref {spectralmeyer}) is:

\begin{equation}
u(x)=\ \sum_m \sum_n \gamma_{mn} f_{m}(2^{-m}x-n).
\label{series}
\end{equation}
The use of the summation (\ref{series}) requires a truncation in $m$ and $n$ that preserves accuracy. The truncation uses the spatial decay of the functions $f_m$ (see \cite{elliott}).  It is convenient to center the summation around the term with the smallest (in magnitude) value of the argument ($2^{-m}x-n$). This can be done by defining for each term $m$ of the outer sum the index $\bar{n}_{m}=\lfloor 2^{-m}x\rfloor$ and shifting indexes in the inner sum to $n'=n-\bar{n}_{m}$. The parameter $m$ governs the scale, the truncation in $m$ defines a frequency range, which can be modified by rescaling the distance $x$ if necessary, so as to obtain a numerically convenient range $-M\leq m\leq 0$. The truncation in $n$ (the number of dilations) defines the region where the support of $f_m$ is concentrated; we suppose that this range lies between a pair of integers -N and N . We can then  write:

\begin{equation}
u(x)=\ \sum_{m=-M}^{0} \sum_{n=-N+1}^{N} \gamma_{m,\bar{n}_{m}+n} f_{m}(2^{-m}x- \lfloor 2^{-m}x\rfloor -n).
\label{field}
\end{equation}
More information can be found in \cite{elliott, cameron}. A technical issue concerning the generation of Gaussian random variables should be pointed out. The number of random variables needed to sample the field scales as $2^{M+1}x$,  and this imposes severe demands on the storage in the computer. To overcome this problem, a reversible pseudorandom number generation is necessary to sample $ \gamma_{mn}$. Simple linear congruential generators with moduli
around $2^{31}$ can exhaust their period in few minutes in a conventional PC, and the resulting poor distribution of the samples can dramatically bias simulation results for sample sizes much smaller than the period length. To overcome this problem,  we used reversible multiple linear congruential generators with many long streams and substreams \cite{Lecuyer} that provide periods  of approximately $2^{191}$ .

We now explain how to modify the construction of the preceding section so as
to impose on the sampled velocity field the non-Gaussian characteristics
observed in the experimental data.

We introduce non-Gaussianity into the numerical experiment via the coefficients $\gamma_m$. In the previous paragraphs the $\gamma_m's$ were Gaussian variables; now we construct a new distribution for the $\gamma_m's$ that preserves the mean and variance of the original Gaussian distribution but with a skewness different from 0. The skewness of a randon variable $\eta$ is $<\eta^3>/<\eta^2>^{3/2}$; it is zero for a Gaussian variable. We describe a transformation that maps a Gaussian variable on a non-Gaussian
variable with the same mean and variance but with a prescribed, non-zero skewness controlled by a single parameter. A Gaussian variable with mean zero and unit variance has the probability density

\begin{equation}
p(x)=\frac{1}{\sqrt{2\pi}} e^{-x^2/2}.
\label{gaus}
\end{equation}

The following change  of variables yields a new skewed probability density function, with a negative skewness as in the cases we consider:

\begin{equation}
y(x)=\frac{-(e^{-ax} -e^{a^2/2})}{\sqrt{(e^{2a^2}-e^{a^2})}}. \label{neg_ng}
\end{equation}

The new probability density function $g(y)$ is obtained by calculating $g(y)=p(x)/|\frac{dy}{dx}|$ provided $y(x)$ is a monotonic function. This formula provides a new distribution with zero mean, unit variance and a skewness controlled by the parameter $a$. For the Eq.(\ref{neg_ng}) the skewness coefficient $c_3$ reads:

\begin{equation}
c_3=\frac{-[(e^{3a^2/2} -e^{a^2/2})^3 + (\sqrt{3}e^{7a^2/4} - \sqrt{3}e^{3a^2/4})] }{\sqrt{(e^{2a^2}-e^{a^2})^3}} \label{neg_skew}.
\end{equation}

\begin{figure}
\begin{centering}
\epsfig{file=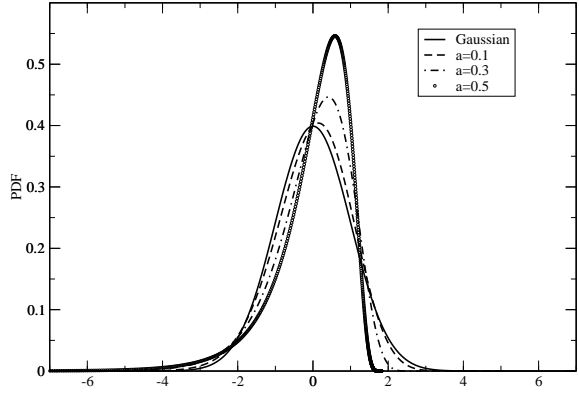, width=7cm,angle=270}
\caption{
Non-Gaussian probability density functions,
obtained by the algorithm in the text
}
\label{1}
\end{centering}
\end{figure}

A plot of these transformed distributions $g(y)$ is presented in Figure 1.%(\ref{1}).

\begin{figure}
\begin{centering}
\epsfig{file=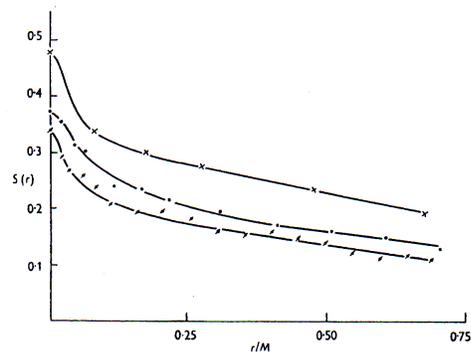, width=14cm,angle=360}
\caption{Skewness as a function of separate $r/M$, from Stewart (\cite{stewart});
$M$ is a reference length in that paper.  Reproduced with permission.}
\label{2}
\end{centering}
\end{figure}

Experimental results for the velocity field were obtained by Stewart in 1951
\cite{stewart}, and are also presented in \cite{batch}, see Figure 2.
The skewness varies with scale, and this is easy to incorporate into our wavelet
representation as the various wavelets describe motion on different scales. The parameter $a(r)$ has been obtained by inversion of the formula (\ref{neg_skew}) for values of $c_3(r)$ corresponding to the experiment for the highest value of $R$ reported in \cite{batch}.

\section{Results}

Having constructed these fields, i.e., written a computer program that samples them,
we proceeded to calculate their properties by Monte Carlo; we now list some of the
results.

1. First, we noticed that if the second order structure has the Kolmogorov form,
and if the field is Gaussian,  then all the structure functions obey the Kolmogorov scaling. This is quite natural and obvious, and was well known
to Kolmogorov (personal communication by G.I. Barenblatt), but we had never
seen it stated in print.
The obvious converses are not necessarily true.

\begin{figure}
\begin{centering}
\epsfig{file=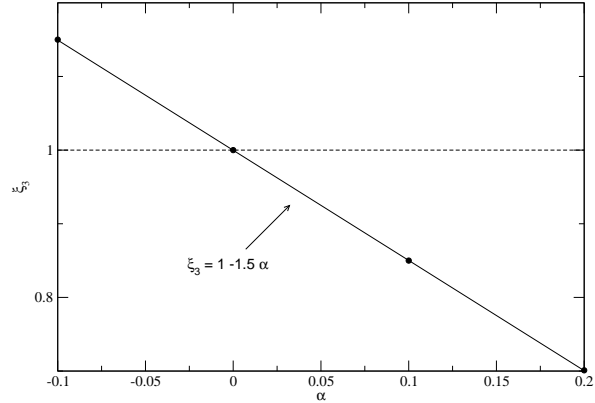, width=7cm,angle=270}
\caption{
Exponent $\xi_3$ of the structure function of order 3
versus $\alpha$ in the relation $E(k)\sim  k^{\frac{-5}{3}+\alpha}$.
The intermittency correction should be where $\xi_3$=1.}
\label{3}
\end{centering}
\end{figure}

2. It is ``almost" a theorem that the third order structure function scales like
$r$ (see \cite{kolm3} as well as\cite{chorinbook,monin}; all one has to assume is an extremely likely bound on the rate of blow-up
of the dissipation as the Reynolds number grows. It is natural to ask then what form must
the second order structure function take if the third order scales as $r$ and
the skewness is as observed. Let the second order structure function scale as $r^{2/3+\alpha}$ and the third order structure
scale as $r^{\xi_3}$; in Figure 3 we plot the computed values of $\xi_3$ as a function of $\alpha$. The relationship between them can be approximated by $\xi_3 =1-\frac{3}{2}\alpha$. The value of $\alpha$ should be the one that yields $\xi_3=1$; thus, within the errors in our sampling scheme, one can construct a velocity field with the second order structure function with the Kolmogorov exponent, the correct third order exponent, and a realistic intermittency; This shows that a correction to the Kolmogorov-Obukhov exponent is not imposed simply by the presence of
intermittency (although we cannot, of course, exclude that it is imposed
by other constraints). Note that this analysis simply confirms that the analysis in Kolmogorov's paper \cite{kolm3}
remains valid in the presence of a realistic skewness. 

3. The next question is maybe the most significant: Suppose the second order structure
function has the Kolmogorov-Obukhov form and the skewness is as observed, what can one say about the higher order structure functions? When we tried to sample these
higher-order functions as above, we found that the variance of the Monte-Carlo
results was very large and did not go down as rapidly as one may expect as the
number of trials went up, leading us to suspect that some moments of the
velocity field diverged.

\begin{figure}
\begin{centering}
\epsfig{file=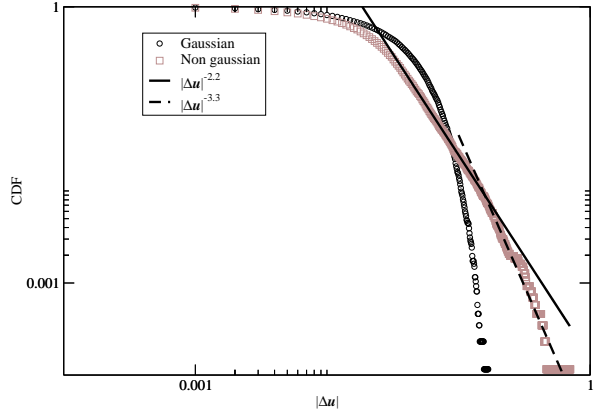, width=7cm,angle=270}
\caption{Cumulative distribution function of $S_1=|\Delta u|$ for a Gaussian and our
non-Gaussian fields at $r/M=1$. M is again a reference length in the tabulated data.}
\label{4}
\end{centering}
\end{figure}

We therefore plotted the data for the variable $x=S_1(r)=|\Delta u|$ at fixed values of  r in log-log scale (see Figure 4).
The data for the non-Gaussian field present a  power-law tail that clearly contrasts with the exponential decay of the Gaussian field. To estimate the form of the decay we construct the cumulative density function (CDF) $P(x>X)$ of the the variable $x=S_1(r)$. We use CDF's rather than the density functions themselves
because they fluctuate less. The behaviour of the CDFs
reveals the convergence, or lack thereof,
of the structure function moments. In the limit as the 
number of samples $n \rightarrow\infty$,
$<S_p>_r=\int (|\Delta u|)^p f(|\Delta u|)d(|\Delta u|)$, where $f$ is the probability density function, the derivative of the CDF.
The CDF for the Gaussian field presents an exponential decay compatible with the finiteness of moments of any order. However the CDF for the velocity differences in the non-Gaussian case
presents a power-law tail whose exponent determines the  order of the moments that are finite. 
The tail of the CDF of $S_1$ scales for small values as a power law with exponent $-2.2\pm0.01$, however, the final part of the CDF, which contain the relevant information about the decay, is well described by an  exponent $-3.3\pm 0.05$ within our experimental error. This means that the PDF will decay approximately as a power law with exponent $-4.3\pm 0.05$ and then produces a threshold for the higher order structure functions to converge at order $p\approx3.3$.
We therefore conclude that, within our model, the structure functions of order $\geq 3.3$ do not exist. In Figure 5 we plotted the CDF's for several values of 
$r$, to show that the behavior we describe is at least approximately independent
of $r$ when $r$ is small.

\begin{figure}
\begin{centering}
\epsfig{file=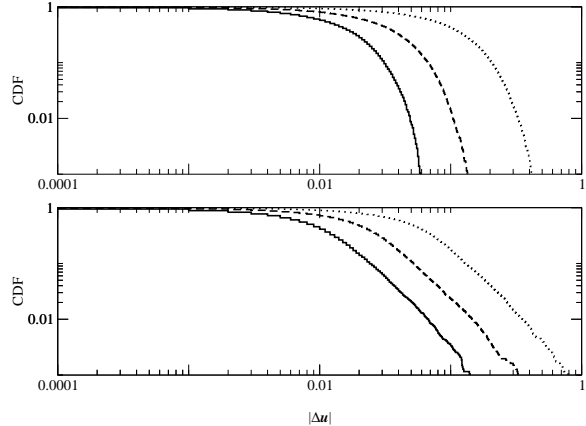, width=7cm,angle=270}
\caption{Cumulative distribution functions  of $S_1$ obtained from a Gaussian field (Top), and a non-Gaussian field, at
several
separations.}
\label{5}
\end{centering}
\end{figure}

This does not of course show that the moments of the true velocity field
in turbulence fail to exist, but only that the data contain a strong enough
deviation from Gaussianity for these moments not to exist.

Finally, we would like to provide a short discussion of sampling errors. All the data presented have been obtained by the least squares method, except the data for the final part of the tail in Figures 4 and 5,
where we have also applied the maximum likelihood method to reduce the estimation error of the exponent \cite{newman}, within  a framework suggested in \cite{tail}. With $95\%$ confidence the errors
are comparable with the thickness of the lines drawn in all cases.  The regions of the sampling  where aliasing problems appear in the FFT have been discarded in the analysis.

\section{Conclusions}

Despite the non-uniqueness of the fields we have constructed, some conclusions
can be drawn from the computations above; in particular:

(1) The original Kolmogorov-Obukhov $-5/3$ spectrum is consistent with
the ``exact" scaling of the third order structure function even in the
presence of intermittency; intermittency does not necessarily require
that the  spectrum be modified. As long as one considers only the lower order
structure functions originally considered by Kolmogorov and Obukhov, our model
provides no reason to modify their original scaling.

(2) The Kolmogorov scaling of the structure functions and its extensions is exact at all orders for Gaussian fields; however, the data for real flows reveal
enough departure from Gaussianity for structure functions of order higher than a threshold larger than three  
to diverge as the Reynolds number grows; if this happens, the 
higher-order exponents become strongly dependent on the Reynolds number, their scaling laws depend on the Reynolds number, and the
proposals of Barenblatt, Chorin, and Prostokishin become eminently
reasonable, as well as those of Mandelbrot and Constantin.
Thus, within the limitations of our model, the complete similarity assumption, on which the
extension of the Kolmogorov scaling rests, fails above the threshold but not below it; when it fails
it should be replaced by an incomplete similarity assumption as in the papers quoted above.

\section{acknowledgements}

We would like to thank Prof. G.I. Barenblatt for inspiration, discussions,
and helpful comments, and Prof. C. Stone for help with the analysis of the data. This work was supported in part by National Science Foundation Grant DMS DMS-0410110 and in part by the Office of Science, Office of Advanced Scientific Computing Research, Mathematical, Information, and Computational Sciences Division, Applied Mathematical Sciences Subprogram of the U.S. Department of Energy, under Contract DE-AC0376SF00098. A.A. also acknowledges financial support by DGES of the Spanish Government Grant No. BFM-2003-08258.

\end{document}